\documentclass{appolb}
\usepackage{epsfig}

\begin{document}
\date{\today}
\pagestyle{plain}
\newcount\eLiNe\eLiNe=\inputlineno\advance\eLiNe by -1
\title{{Mean first passage time for a Markovian jumping process} }
\author{ A. Kami\'nska and T. Srokowski
\address{Institute of Nuclear Physics, Polish Academy of Sciences, PL -- 31-342 Krak\'ow,
Poland }}

 \maketitle

\begin{abstract}
We consider a Markovian jumping process with two absorbing barriers,
for which the waiting-time distribution involves a
position-dependent coefficient. We solve the Fokker-Planck equation
with boundary conditions and calculate the mean first passage time
(MFPT) which appears always finite, also for the subdiffusive case.
Then, for the case of the jumping-size distribution in form of the
L\'evy distribution, we determine the probability density
distributions and MFPT by means of numerical simulations. Dependence
of the results on process parameters, as well as on the L\'evy
distribution width, is discussed.
\end{abstract}


\maketitle
\section{Introduction}

Transport processes in physical systems are usually considered in
the diffusion limit of large distance at long time. In realistic
situations, however, the available space is finite and this must be
taken into account. In the framework of the stochastic description,
that restriction means that the system possesses an absorbing and/or
reflecting barrier at which the probability distribution vanishes.
Therefore, the corresponding equations must involve boundary
conditions. Effects connected with the final size of the system are
especially pronounced for small objects and they are encountered in
many applications of the stochastic processes to systems of high
complexity. Taking into account the absorbing barriers is essential
in dealing with several population and environmental problems. For
example, models of species extinction involve the size of a refuge
in which life conditions are favorable, whereas they are extremely
harsh outside \cite{esc}. Inclusion of size of the refuge is also
crucial in a model of the infection of Hantavirus in the deer mouse,
based on biological observations in North America \cite{abr}.

Time characteristics of the escaping process involve the first
passage time density distribution, defined as a probability that the
time the particle needs to reach the absorbing boundary is within
the interval $(t,t+dt)$, provided the particle was initially at a
given point $x_0$ \cite{ris}. The average of that distribution --
the mean first passage time (MFPT)  -- is a useful quantity to
estimate the speed of transport for systems which are defined on the
restricted area. The MFPT can be calculated also for the boundless
systems, if a potential restricts the domain.

For example, an application of the diffusion problem
on the finite interval to the heat
conduction between hot and cold baths is recently of wide interest.
There are several attempts to link such thermodynamical phenomena as
heat conductivity to the dynamical diffusion. In this context, the
problem of validity of Fourier's law, as a counterpart to the Fick
diffusion law, is especially interesting. Applicability of dynamical
processes, which are characterized by the anomalous diffusion
coefficient, became clear when the anomalous heat
conductivity in classical one-dimensional lattice systems has been found
\cite{lep1}. A model, called ``dynamical heat channels'' \cite{den,cip},
can be constructed by introduction some simplifications,
e.g. by neglecting the interactions between particles. Then the dynamics
can be handled by the decoupled CTRW which implies all kinds of the
anomalous diffusion. In this particular model, the subdiffusive case
requires long tails of the waiting time distribution;
as a result the average waiting time, as well as MFPT \cite{yus},
is infinite. For the heat conduction process that would mean a perfect insulator.

However, that uncoupled version of CTRW does not take into account that,
in general, the system may be inhomogeneous, i.e. its
parameters depend explicitly on the spatial variable.
This happens in the complex systems where long-range space correlations are
important and the medium structure is crucial for the system properties. As an
example can serve the transport on the fractal objects \cite{met3,met4,tar} and, since
fractals are ubiquitous in nature, its numerous manifestations in various
branches of science. The transport coefficients must vary with the position if
one describes the dynamical properties of materials containing impurities
and defects. Physical problems which are considered in this context involve
conductivity of amorphous materials, the ionic conductors, dynamics of
dislocations, transport of a dye in porous materials (quenched disordered media)
\cite{bou}.
In the case of the heat conduction, deviations of model calculations
from Fourier's law indicate that the asymptotic temperature gradient is
nonuniform and they point at long-range effects \cite{lep}. The MFPT for
a process which can correspond e.g. to the Langevin equation with the
multiplicative noise, and which is also described by the Fokker-Planck
equation with the variable coefficient, was calculated in Ref. \cite{kwo}.

In the present paper we evaluate the MFPT for a jumping process which is
a version of the CTRW: it is Markovian and takes into account the
spatial dependences of the problem by introduction the $x$-dependent waiting time
distribution.

Effects connected with the finite size of the system are especially
pronounced if the particle performs long jumps, namely for the
L\'evy flights, when the second moment of the probability
distribution is infinite. If the distance between boundaries is
small, compared to width of the jump length distribution, the tails
hardly influence the dynamics and the essence of the L\'evy process
remains hidden. On the other hand, presence of the barriers makes
all the moments convergent.

The paper is organized as follows. In Sec.II we present the definition
and main properties of the jumping process. In Sec.III the MFPT for the
system which possesses two absorbing barriers is calculated. The
consequences of introduction of distributions with long tails (L\'evy
flights) is analysed in Sec.IV and dependence on the process parameters
is discussed. The main results are summarized in Sec.V.

\section{Description of the process}

The process we consider in this paper is a step-wise one-dimensional
Markov process defined
in terms of the jumping size distribution $Q(x)$ and the Poissonian
waiting time distribution
\begin{equation}
\label{poi}
P_P(t)=\nu(x){\mbox e}^{-\nu(x)t},
\end{equation}
where $\nu(x)$ is the jumping rate \cite{kam}. The process value $x(t)$
is constant between consecutive jumps. Since $\nu$ depends on $x$,
the process is a generalization of the usual, uncoupled CTRW. The master
equation is the following
\begin{equation}
\label{ma} \frac{\partial p(x,t)}{\partial
t}=-\nu(x)p(x,t)+\int_{-\infty}^\infty Q(x-x')\nu(x')p(x',t)dx'.
\end{equation}
In the following, we assume
the scaling form, $\nu(x)=|x|^{-\theta}~~(\theta>-1)$, for $\nu(x)$
which was applied e.g. to study
the diffusion on fractal objects \cite{osh}. Moreover, it was used to describe
the transport of fast electrons in a hot plasma \cite{ved} and the
turbulent two-particle diffusion \cite{fuj}.

A natural choice for the distribution $Q(x)$ is the Gaussian:
\begin{equation}
\label{gau}
Q(x)=\frac{1}{\sigma\sqrt{2\pi}} {\mbox e}^{-x^2/2\sigma^2}.
\end{equation}
The corresponding master equation for the jumping process
can then be approximated -- by means of the Kramers-Moyal expansion -- by the following
Fokker-Planck equation \cite{sro}
 \begin{equation}
  \label{1}
  \frac{\partial p(x,t)}{\partial t} = \frac{\sigma^2}{2}
  \frac{\partial^2 [|x|^{-\theta}p(x,t)]}{\partial x^2},
  \end{equation}
where $\sigma$ is the width of the distribution $Q(x)$.
The solution, with the initial condition $p(x,0)=\delta(x)$, is given by
\begin{equation}
   \label{rozw}
   p(x,t)=C_{\theta} \frac{|x|^{\theta}\exp(-\frac{2|x|^{2+\theta}}
{\sigma^2
(2+\theta)^2t})}{(\sigma^2t/2)^{\frac{1+\theta}{2+\theta}}},
   \end{equation}
where $C_{\theta}=\frac{1}{2 \Gamma(\frac{1+\theta}{2+\theta})}
|2+\theta|^{\frac{\theta}{2+\theta}}$. The mean squared displacement
can be directly evaluated: $\langle x^2(t) \rangle \sim
t^{\frac{2}{2+\theta}}$. Then for $\theta\in(-1,0)$ the superdiffusion emerges,
for $\theta>0$ we get
the subdiffusion. The normal diffusion takes place for
$\theta=0$. Therefore, this Markovian process involves all kinds of diffusion.

The other form of $Q(x)$ is the L\'evy distribution which is also
stable and has the broad, power-law tails
$|x|^{-\mu-1}~~(0<\mu<2$)\cite{VMZ}. The Kramers-Moyal
approximation of the master equation (\ref{ma}) produces in this
case the following fractional equation:
\begin{equation}
\label{frace} \frac{\partial p(x,t)}{\partial
t}=K^\mu\frac{\partial^\mu[|x|^{-\theta}p(x,t)]}{\partial|x|^\mu},
\end{equation}
instead of the Fokker-Planck equation (\ref{1}).
The solution of the Eq. (\ref{frace}) represents the L\'evy process and it can be
expressed in terms of the Fox function in the following form \cite{sro,sro1}
\begin{eqnarray}
\label{solp0}
p(x,t)=\frac{a}{\mu}H_{2,2}^{1,1}\left[a|x|\left|\begin{array}{l}
(1-1/\mu,1/\mu),(1/2,1/2)\\
\\
(0,1),(1/2,1/2)
\end{array}\right.\right].
\end{eqnarray}
where $a\sim t^{-1/(\mu+\theta)}$. The solution (\ref{solp0}) is
correct, and equivalent to the solution of the master equation (\ref{ma}),
in the diffusion limit of large both $x$ and $t$. Since all moments
of order $\delta\ge\mu$ of the distribution $p(x,t)$, Eq.(\ref{solp0}),
are divergent, the kind of diffusion process cannot be determined from
time dependence of the second moment.
Instead, one can introduce fractional moments of the order
$\delta<\mu$. Alternatively, the renormalized moment of the order $\mu$
\cite{sro} allows us to characterize the diffusion properties of the
system in the same way as for the Gaussian case and to
distinguish the normal diffusion ($\theta=0$), subdiffusion ($\theta>0$)
and the superdiffusion ($\theta<0$).

Presence of the absorbing barriers must modify the probability
distribution $p(x,t)$ for both choices of $Q(x)$: the distributions
dwindle with time due to the absorption and the broad tails in the
L\'evy case are cut off. As a consequence, all the moments are
finite. We discuss those problems in the next sections.

 \section{Fokker-Planck equation with the boundary conditions}

We consider a one-dimensional motion which is restricted to an
interval $[0,L]$. The particle performs jumps defined by the
probability distributions $P_P(t)$ and $Q(x)$ according to the
Eqs.(\ref{poi}) and (\ref{gau}). The end points of the interval, 0
and $L$, are regarded as the absorbing barriers; the probability
distribution is given by the Fokker-Planck equation (\ref{1}) with
the initial condition $p(x,t=0)=\delta(x-x_0)$ $(0<x_0<L)$ and with
the following boundary conditions
\begin{equation}
  \label{bc}
p(0,t)=p(L,t)=0.
\end{equation}
Eq. (\ref{1}) for this problem can be solved by separation of
the variables. Let us assume the particular solution in the form
$p(x,t)=\phi(t)\psi(x)$. Inserting this
ansatz to the Eq.(\ref{1}) yields two equations; the function $\phi$
can be easily determined: $\phi(t)=C\exp(-\lambda^2t)$, where
$\lambda=$const. For the function $\psi$ we get the equation
\begin{equation}
  \label{4}
  \frac{\partial^2 y(x)}{\partial
  x^2}+\frac{2\lambda^2}{\sigma^2}x^\theta y(x)=0,
 \end{equation}
where $y(x)=x^{-\theta}\psi(x)$.
The solution of the Eq.(\ref{4}) can be expressed in terms of
the Bessel functions $J_\nu(x)$ in the following form \cite{kamk}
\begin{equation}
  \label{5}
  y(x)=A\sqrt{x}J_{1/(\theta+2)}\left(\frac{2\sqrt{2}\lambda}
  {\sigma(\theta+2)}x^{(\theta+2)/2}\right)
\end{equation}
which satisfies the condition $\psi(0)=0$. The second boundary
condition, $\psi(L)=0$, allows us to determine the parameter
$\lambda\equiv\lambda_n$ by means of the zeros of the Bessel function $\gamma_n$:
\begin{equation}
  \label{7}
\lambda_n=\frac{\sigma(\theta+2)}{2\sqrt{2}}\frac{\gamma_n}{L^{(\theta+2)/2}}.
\end{equation}
The general solution can be obtained by summing up over all values
of $\lambda_n$:
\begin{equation}
  \label{8}
  p(x,t)=x^{\theta+1/2}\sum_n A_n J_{1/(\theta+2)}\left(\frac{\gamma_n}
  {L^{(\theta+2)/2}}x^{(\theta+2)/2}\right)
  \exp\left(-(\frac{\sigma(\theta+2)}{2\sqrt{2}}\frac{\gamma_n}{L^{(\theta+2)/2}})^2 t\right).
\end{equation}
The form of the constant $A_n$ follows from the initial condition. The orthogonality
property of the Bessel function produces, after some algebra, the following expression:
\begin{equation}
  \label{11}
  A_n=\frac{\theta+2}{L^{\theta+2}}\frac{\sqrt{x_0}J_{1/(\theta+2)}\left(\frac{\gamma_n}
  {L^{(\theta+2)/2}}x_0^{(\theta+2)/2}\right)}{[J'_{1/(\theta+2)}(\gamma_n)]^2}.
\end{equation}

The series representation of the distribution $p(x,t)$, Eq. (\ref{8}),
is convergent for all $x$ and $t$. An example of the time evolution of $p(x,t)$,
calculated according to the Eq.(\ref{8}) in which 40 terms has been taken into
account, is presented in Fig.1. The distributions shift to the right with time
and their normalization integral becomes smaller -- due to absorption at the
boundary $x=L=5$.
\begin{figure}[h]
\includegraphics[angle=0,width=0.7\textwidth]{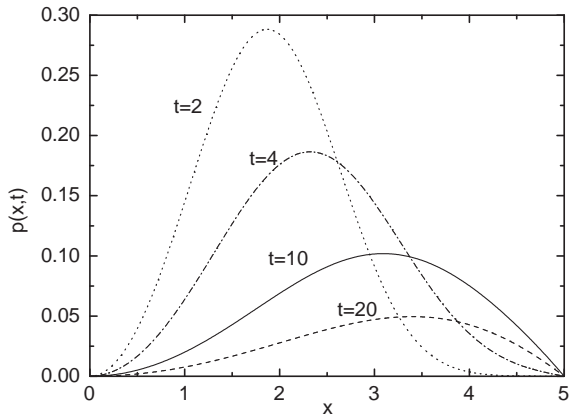}
\caption{The solutions of Eq.(\ref{1}) with boundary conditions
(\ref{bc}) for $\theta=1$.}
\end{figure}

Having the distribution $p(x,t)$ calculated, we can determine
the survival probability: the probability
that the particle is still inside the interval $(0,L)$, i.e. it has not yet reach the
absorbing barrier. It can be obtained by means of the formula
$S(t)=\int_0^Lp(x,t)dx$ and it determines the first passage time density
distribution $f(t)=-dS(t)/dt$. The averaging over that distribution produces the MFPT:
\begin{equation}
  \label{12}
  T=\int_0^\infty tf(t)dt=\int_0^Ldx\int_0^\infty p(x,t)dt.
\end{equation}
In the case of our jumping process, the direct
evaluation of the integral yields
\begin{equation}
  \label{st}
S(t)=\frac{2x_0}{\sqrt{L}}\sum_n\frac{J_{1/(\theta+2)}\left(\frac{\gamma_n}
  {L^{(\theta+2)/2}}x_0^{(\theta+2)/2}\right)}{
\gamma_n J_{-(\theta+1)/(\theta+2)}(\gamma_n)}
\exp\left(-(\frac{\sigma(\theta+2)}{2\sqrt{2}}
  \frac{\gamma_n}{L^{(\theta+2)/2}})^2 t\right)
\end{equation}
where we utilized simple properties of the Bessel function. To obtain the MFPT
we need to integrate $S$ over time:
\begin{equation}
  \label{13}
T=\int_0^\infty
S(t)dt=\frac{8x_0L^{\theta+3/2}}{\sigma^2(\theta+2)}
 \sum_{n}\frac{J_{1/\theta+2}\left(\frac{\gamma_n}
  {L^{(\theta+2)/2}}x_0^{(\theta+2)/2}\right)}
  {[J'_{1/\theta+2}(\gamma_n)]\gamma_n^3}.
\end{equation}

Fig. 2 presents the survival probability for some values of
$\theta$, both positive and negative, calculated from Eq.(\ref{st}).
For the problem without absorbing barriers, the case with $\theta<0$
corresponds to the superdiffusion, whereas with $\theta>0$ -- to the
subdiffusion. The figure shows that the tails are always
exponential. Moreover, $S(t)$ rises with $\theta$ for any $t$, as
expected. For large values of $\theta$, beginning of the curve is
flat which means that trajectories can hardly escape at short time
due to the strong trapping. Since $S(t)$ becomes actually
exponential, the MFPT is finite for all $\theta$ (see
Eq.(\ref{13})), also for those which correspond to the subdiffusion.
This result is in contrast to that of the decoupled CTRW which
predicts divergence of the MFPT in the subdiffusive case \cite{yus}.
More precisely, the decoupled CTRW in the subdiffusive case is
non-Markovian and it assumes the waiting time distribution in the
power law form. Then the mean time of a single jump is infinite. For
the problem with the absorbing barriers, one can derive a formula
for MFPT directly from the waiting time distribution and the MFPT
appears infinite. For the process presented in this paper, the
waiting time distribution is exponential and the subdiffusion
results from the $x-$dependence of its coefficient, i.e. from
nonhomogeneity of the medium. Introduction of that dependence has
important physical implications. As regards the application to the
heat conduction problem, the subdiffusive thermal conductivity
becomes possible also for the systems which are not perfect thermal
insulators and then more realistic.
\begin{figure}[h]
\includegraphics[angle=0,width=0.7\textwidth]{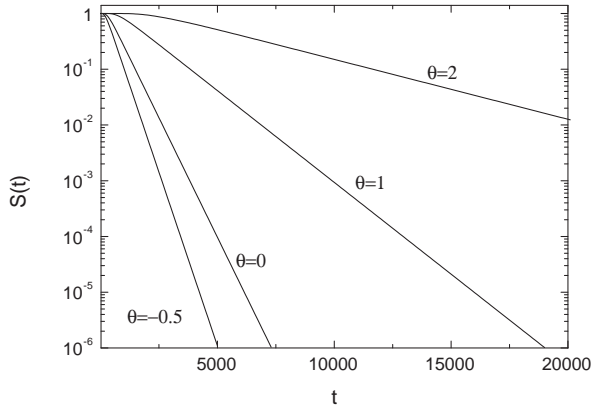}
\caption{The survival probability calculated from Eq.(\ref{st}) for
the values of $\theta$ which corresponds to the superdiffusion
($\theta<0$), normal diffusion ($\theta=0$), and subdiffusion
($\theta>0$).}
\end{figure}

\section{L\'evy flights between the absorbing barriers}

In this section we analyse the jumping process for the system
restricted by two absorbing barriers for which the jumping size
distribution is given by the L\'evy distribution. We calculate the
probability density distributions and the MFPT as a function of both
the L\'evy index $\mu$ and the parameter $\theta$.

The MFPT problem for the L\'evy flights on the bounded domain, both
with and without a potential, is studied extensively in recent years;
beside the MFPT, the first passage time distribution has been evaluated
as a function of the parameters of the L\'evy distribution,
which, in general, can be asymmetric.
Since the analytical approach is very difficult in this case,
most of the studies rely on the Monte Carlo simulations \cite{DGH,FZ,Q}.
Nevertheless, recently an analytical solution to the fractional equation
with the boundary conditions, which describes the L\'evy flights
in a homogeneous medium ($\theta=0$), has been found \cite{ZK}.

 The L\'evy
distribution represents the general stable distribution and in that
sense it is a generalization of the Gaussian. It accounts for
processes for which the second moment of the probability density
distribution diverges: the standard central limit theorem does not
apply in this case. Phenomena which exhibit distributions with long
tails are frequently encountered in nature. They are typical for
systems of high complexity, in particular biological \cite{wes},
social, and financial ones. Therefore, the theory of the L\'evy
flights is widely applicable to problems from various branches of
science and technology.

One can expect that presence of the barriers will influence the
stochastic dynamics particularly strong in the case of the L\'evy
processes. If the interval length $L$ is small compared to the width
parameter $\sigma$ of the jump length distribution, the power-law
tails of the distribution will not manifest themselves. In
particular, all moments become finite.

We assume the jump length distribution in the form
\begin{equation}
\label{lev}
Q(x)=\sqrt{2/\pi} \int_0^\infty \exp(-\sigma^\mu k^\mu)\cos(kx)dk,
\end{equation}
as well as the Poissonian waiting time distribution (\ref{poi}). We
determine the probability density distribution $p(x,t)$ by means of
the Monte Carlo method. The L\'evy-distributed jump-size density has
been generated by using the algorithm from Ref. \cite{S}. The time
evolution of individual trajectories, which start with the same
initial condition, has been performed by sampling
consecutive values of the jumping size and the waiting time interval
from the densities $Q(x)$ and $P_P(t)$, respectively. The final results
have been obtained by averaging over those individual
trajectories. Fig.3 presents
the time evolution of $p(x,t)$ for both positive and negative
$\theta$. Similarly as in the case of the Fokker-Planck equation,
the distributions terminate abruptly at the barrier position and
they shrink with time due to the absorption. The initial delta
function at $x=2$ is visible up to a long time.

\begin{figure}[h]
\includegraphics[width=0.7\textwidth]{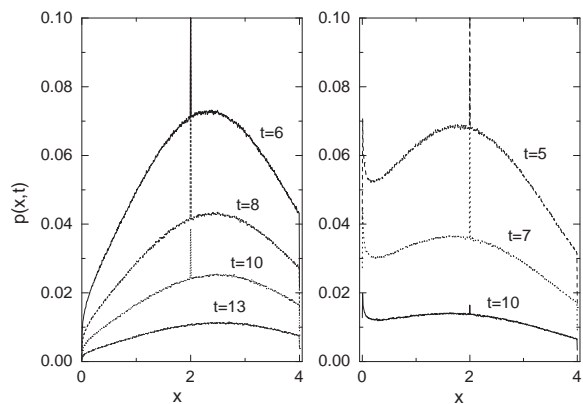}
\caption{The time evolution of probability density distribution for
$Q(x)$ in the form of the L\'evy distribution (\ref{lev}) with the
parameters $\mu=1.5$ and $\sigma=1$, calculated for two values of
$\theta$: $\theta=0.2$ (left part) and $\theta=-0.1$ (right part).
The vertical lines near the absorbing barrier $x=4$ represent very
fast fall of the distributions due to absorption.}
\end{figure}
In the case of the process presented
in Fig.3 the width parameter $\sigma$ of the driving distribution $Q(x)$ is
large, compared to the interval size $L$, and the results are not sensitive to
its tails. On the other hand, in the limit of large
$\bar L=L/\sigma$ one can expect that
$p(x,t)$ converges to the distribution which corresponds to the process
without absorbing barriers. In this case, the resulting distribution
should not depend on $\sigma$ and the asymptotics should be completely
determined by $\mu$. Indeed, Fig.4 demonstrates that for
$\sigma=0.003$ the tail approaches the form $\sim x^{-1-\mu}$, before
it is cut abruptly at the barrier position. For the slightly larger value
of this parameter, $\sigma=0.004$, the power-law asymptotics fails to appear.
The tails of $p(x,t)$ become $\sigma$-dependent for relatively large $\sigma$
because the importance of the tails of $Q(x)$ gradually declines with $\sigma$.
\begin{figure}[h]
\includegraphics [width=0.7\textwidth]{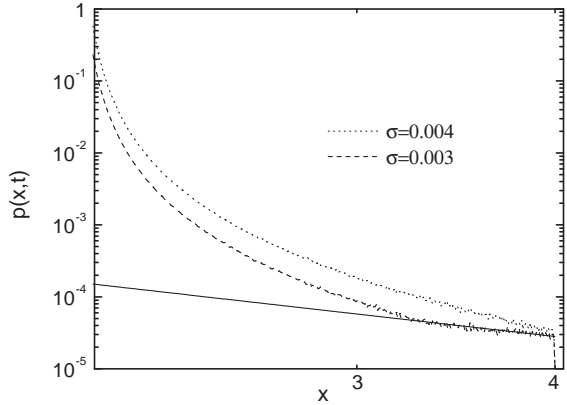}
\caption{The log-log plot of the tails of probability density
distribution $p(x,t)$, calculated with $\theta=0.2$ and $\mu=1.5$,
for two values of the width parameter $\sigma$. The solid line
denotes the function $\sim x^{-5/2}$.}
\end{figure}

The same procedure allows us to determine MFPT: the average time $T$
the edge of the interval, i.e. either 0 or $L$, is reached. The dependence
of MFPT on both $\mu$ and $\sigma$ is presented in Fig.5 for the case
$\theta=0$. In general, $T$ rises
with $\mu$ because then the probability of large jumps falls.
Since that effect results from the power-law
tails, it is weak (the curves are flat) if only the central part
of the distribution -- similar
for all $\mu$ -- is involved, i.e. for large $\sigma$. On the other hand,
small values of $\sigma$ result
in a strong dependence $T(\mu)$, which becomes exponential for $\sigma=0.003$.
This shape persists for even smaller $\sigma$.
\begin{figure}[h]
\includegraphics[width=0.7\textwidth]{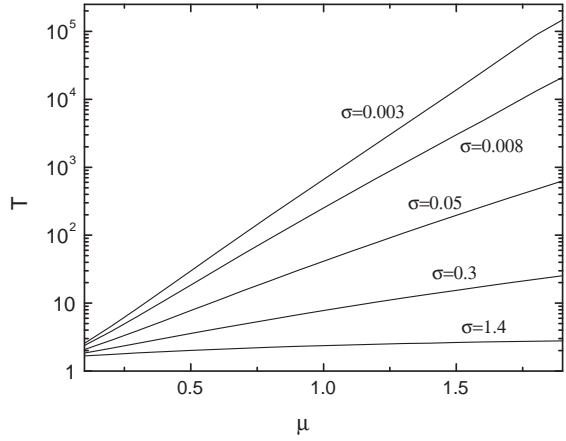}
\caption{The MFPT as a function of the parameter $\mu$ calculated for $\theta=0$
and some values of $\sigma$.}
\end{figure}

The parameter $\theta$ is crucial to the speed of the transport; in
absence of the absorbing barriers, the cases with large values of
$\theta$ correspond to the slow diffusion. It happens because the
traps at large distances becomes more effective when $\theta$
increases and the transport is hampered due to the long waiting
times. Consequently, one can expect that the MFPT will rise with
$\theta$. This conclusion is illustrated in Fig.6. Growth of the
function $T(\theta)$ is especially pronounced for large values of
$\mu$ (close to Gaussian case $\mu=2$) because then the average jump
length is relatively small and a large number of jumps is needed to
reach the barrier. Transition from the negative to positive values
of $\theta$ -- which corresponds to the change of kind of diffusion
in the problem without absorbing barriers -- is smooth for all
$\mu$. The latter conclusion holds also for $\mu=2$, according to
the Eq.(\ref{13}).
\begin{figure}[h]
\includegraphics[width=0.7\textwidth]{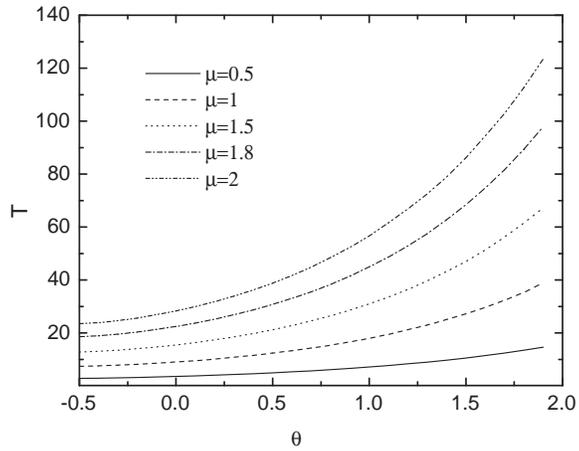}
\caption{The MFPT as a function of the parameter $\theta$,
calculated with $\sigma=0.3$, for some values of $\mu$.}
\end{figure}

\section{Summary and discussion}

We have analysed the one-dimensional, step-wise jumping process which is
defined on the finite interval, bounded by two
absorbing barriers. Since the waiting
time distribution involves the variable, position-dependent coefficient,
the inhomogeneity of the medium has been taken into account. That,
power-law, dependence is a reason of the anomalous behaviour: the
diffusion can be both weaker and stronger than normal. The calculated
MFPT is large for the subdiffusion and rises with the parameter $\theta$,
nevertheless it always assumes a
finite value. Physical reason behind the latter outcome -- which
can never be concluded from the traditional, uncoupled CTRW models
of the subdiffusion -- is the weakening of trapping with the increased
distance and, as a result, the effective mean waiting time is finite.
From the point of view of modelling of the anomalous heat conduction,
the introduction of the $x$-dependent waiting time distribution
allows us to describe the subdiffusive heat transport for realistic
systems: for those which are not perfect insulators.

Restrictions imposed on the system by the existence of the absorbing
barriers are pronounced if we allow for long jumps, i.e. if the jumping
size distribution $Q(x)$ is of the L\'evy form. The power law tails of
that distribution can influence the probability density distribution
of the process, $p(x,t)$, only if $Q(x)$ is narrow,
compared to the distance between the barriers, i.e. to the system size.
In that limit, the sections of $p(x,t)$ which are close to the barrier
assume the power law shape in the same form as the tails for the problem
without the barriers. Conversely, for the broad $Q(x)$ those tails are hardly
visible and the MFPT is almost independent of the L\'evy parameter $\mu$.
The dependence of MFPT on the parameter $\theta$ is similar to that
for the Gaussian $Q(x)$: large values of $\theta$, for which the transport
is strongly hampered by the traps, result in large $T$. Nevertheless,
it remains finite, for any $\mu$ and $\theta$.

\end{document}